**Title: "Autonomous Fabrication of Tailored Defect Structures in 2D Materials using Machine Learning -enabled Scanning Transmission Electron Microscopy"**

**Authors:** Zijie Wu[1], Kevin M. Roccapriore[1,2], Ayana Ghosh[3], Kai Xiao[1], Raymond R. Unocic[4], Stephen Jesse[1], Rama Vasudevan[1], Matthew G. Boebinger[1*]

[1] Center for Nanophase Materials Sciences, Oak Ridge National Laboratory, Oak Ridge, TN, USA

[2] AtomQ, Knoxville, TN, USA

[3] Computational Science and Engineering Division, Oak Ridge National Laboratory, Oak Ridge, TN, USA

[4] Department of Materials Science and Engineering, North Carolina State University, Raleigh, NC, USA

*Corresponding author boebingermg@ornl.gov

**Abstract:**

*Materials with tailored quantum properties can be engineered from atomic scale assembly techniques, but existing methods often lack the agility and accuracy to precisely and intelligently control the manufacturing process. Here we demonstrate a fully autonomous approach for fabricating atomic-level defects using electron beams in scanning transmission electron microscopy (STEM) that combines advanced machine learning and automated beam control. As a proof of concept, we achieved controlled fabrication of MoS-nanowire (MoS-NW) edge structures by iterative and targeted exposure of $MoS_2$ monolayer to a focused electron beam to selectively eject sulfur atoms, utilizing high-angle annular dark-field (HAADF) imaging for feedback-controlled monitoring structural evolution of defects. A machine learning framework combining a random forest model and convolutional neural networks (CNN) was developed to decode the HAADF image and accurately identify atomic positions and species. This atomic-level information was then integrated into an autonomous decision-making platform, which applied predefined fabrication strategies to instruct beam control about atomic sites to be ejected. The selected sites were subsequently exposed to localized electron beam using an FPGA-controlled scan routine with precise control over beam positioning and duration. While the MoS-NW edge structures produced exhibit promising mechanical and electronic properties,[1-3] the proposed autonomous fabrication framework is material-agnostic and can be extended to other 2D materials for the creation of diverse defect structures and heterostructures beyond $MoS_2$.*

**Introduction:**

Designing materials with specific functional properties requires precise control over material morphology and chemistry, often beyond the capabilities of conventional manufacturing techniques. Atomically precise fabrication has emerged as a class of next generation approaches, offering remarkable control over a material's structure and properties down to the atomic scale.[4-7] Such a level of precision is critical for driving breakthroughs in various fields, including quantum sensing [8, 9] and communications [10, 11]. With advancements in aberration-corrected scanning transmission electron microscopy, the electron beam can be used as a tool for direct atomic-scale fabrication.[4, 12, 13] Atomic manipulation utilizing the STEM electron beam has been demonstrated in several material classes in predominantly 2D materials such as graphene and transition metal dichalcogenides (TMDs).[14-20] Earlier studies by Susi and coworkers [21-23] and Dyck et al.[24] have demonstrated the possibility of performing controlled single atom manipulation of Si defects in monolayer graphene through manually directed electron beam. Dyck et al. additionally demonstrated the capability to perform single dopant control with a variety of other atomic species.[25-27] Recent work by Schwarzer et al.[28] also showcased the possibility of using neural network to model and drive Si migration in pristine graphene. However, many of these early efforts used primarily manual control over the experiments, relying on the user to provide the strategy and control the beam position for structure manipulations and microscope operation. To avoid the limitations of this manual method of fabrication, automatic feedback control electron microscope experimentation has been developed in recent years.[13, 29, 30] One such system developed by Jesse et al. utilized both the Annular Dark Field (ADF) scan signal intensity and the fast Fourier transform of the signal for feedback and to control the placement of dopant atoms, milling of nanopores, or crystallization of scanned areas for automated experiments.[13, 31]

Previous works conducted by our group have utilized a feedback-controlled beam control system to form different defect structures within the $MoS_2$ TMD material system using 60 kV electron beam irradiation.[32, 33] Electron beam irradiation was found to be capable of ejecting S atoms from the lattice to reliably form a variety of defect structures. Previous works have also demonstrated control over the atomic fabrication of line defects formed by rows of single sulfur vacancies, termed single vacancy line (SVL) defects that have theoretically altered local electronic band structure when grown together.[33] Additionally, it has been shown that more complex defect structures such as 1D-2D edge heterostructures along nanopores within the $MoS_2$ monolayer can also be fabricated.[32, 34] These defect structures consist of a metallic $Mo_6S_6$ nanowire structure (MoS-NW) attached to the semiconducting $MoS_2$ lattice with the local properties being determined by the on-lattice edge bonded to the MoS-NW. Through controlling the beam scanning shape and pathways along the crystallographic directions different MoS-NW structures were formed.[32] However, these fabrications relied on favorable statistics that the S atoms would be ejected from the lattice during standard raster scanning, even when rastered over crystallographic directions. So, while these previous studies demonstrated the feasibility of automated electron beam induced fabrication, to increase the precision of this atomic species defect engineering, an alternative approach is needed.

We have explored the incorporation of machine learning in atomic fabrication by developing a more generalized approach to autonomous electron beam fabrication through the utilization of machine learning for *in situ* atom detection, [33, 35-37] where a fully autonomous electron beam fabrication experimental platform was developed through an ensemble learning and iterative training (ELIT)-based approach and additional machine learning based postprocessing for real-time analysis of the data flow in STEM. [33] ELIT was shown to avoid previously observed problems relying on single models constructed from simulated training data, producing a more robust atom identification/classification procedure capable of overcoming these limitations[35] and ultimately improving confidence in identifying atomic structures. ELIT-based atom sites classification have also been paired with electron energy loss spectroscopy (EELS) to monitor beam-induced structural evolution and guide spectral measurement at regions of interest.[36] ELIT has the potential to dramatically improve the throughput of the STEM experiments to satisfy the requirement of rapid atomic fabrication workflow, since images can be acquired at a faster frame rate and lower electron dose, with classification accuracy still maintained by robust atomic feature identification.[38] However, to further improve upon the previously developed workflow to enable fully autonomous operation for complex defect fabrication, we need a method to autonomously determine where to place the beam to fabricate desired defect based on the location and size of atoms and mesoscale features (e.g. the SVLs and MoS-NWs) in the current state. This method in turn necessitates a method to automatically recognize and label these relevant features in the field of view reliably and rapidly, even as the imaged area begins to deform in a variety of ways. While our previously developed ELIT framework can decode the STEM images to identify and classify atomic species, the ELIT models are based on pretraining models on large amounts of simulated data (and further augmented with a smaller amount of experimental data), making the method less reliable when dealing with kinetically formed features such as SVLs and MoS-NWs that are hard to reproduce in simulations. For these features, we often had to resort to heuristic-based classification criteria that were tedious to implement, challenging to tune, and provide inconsistent results. Recently, there have been multiple studies using machine learning methods for STEM and atomic resolution TEM image segmentation.[39-47] Many of these ML methods effectively treat their tasks as a computer vision method, deploying ML models to directly segment the boundaries of features. However, we argue that there are several key differences in training ML models to classify features in STEM from training "canonical" computer vision models to identify daily objects such as pedestrians or trees:

(1) The domain of atomically resolved STEM images for condensed matter, usually consisting of clusters of high intensities (atoms) distributed on a relatively uniform background, is much smaller than the natural macroscopic images the computer vision models are originally developed for and excel at. As a result, traditional segmentation models are unnecessarily complex and cumbersome to deploy for STEM images and often struggle to produce significantly improved performance and/or generalizability compared to simplistic filtering methods or minimally trained neural networks[35].

(2) It is significantly more expensive to generate enough labeled training data, either through simulations or by manual labeling of experimental data, to train the computer vision models specifically for a STEM-related purpose. Even if such models are trained or transfer learning is attempted on open-source baseline computer vision models (e.g. ImageNet, YOLO), they often suffer from poor transferability (from one material to another, from simulation to experiments) and require significant task-specific tuning.

Collectively, these two points of difference suggest that we can potentially better address feature classification in STEM using a more "small-data" approach – simplistic models built and trained with fewer trainable parameters and hyper parameters, smaller training sets, and heavier feature engineering based on our intuition as material scientists. Spurgeon and coworkers used few-shot learning to segment local patches in STEM images into classes representing multiple microstructures.[43] In this work, we combine a similar few-shot philosophy with ELIT-based atom localization to determine atomically resolved features of interest in STEM images and guide the electron beam positions for targeted defect fabrication. By combining the previously found experimental parameters of $MoS_2$ functional defect generation with this autonomous electron microscopy platform, a workflow was developed to automate the fabrication process and dramatically increase the level of control over the fabrication of the MoS-NW edge structure. This level of control enabled the selective formation of the MoS-NW on pre-defined nanopore edges, free-standing MoS-NWs, as well as the growth of these structures beyond their initial formation. Additionally, during these experiments, we also include the possibility (but not necessity) to use FPGA instead of the native microscope controlling software for advanced and more precise control of beam path and dwell time.

**Results and Discussion:**

The central goal of this workflow is to use the robust autonomous models in our machine learning framework to convert the data stream from the STEM detectors (specifically, the HAADF detector) in real-time into atomic identities and coordinates that can then be used for feedback for fabrication. Using these pieces of information, the STEM scan coils can then be given commands to target specific sites for fabrication of desired morphology. This procedure was shown in this group's previous work by Roccapriore et al.[33]. The current study reports on more advanced decoding and fabrication workflows that are capable of reliably producing more complex defect structures within MoS$_2$ material system. The cyclical workflow for the fabrication of the MoS-NW decorated nanopores can be seen in Figure 1, with the decoding and labeling of the annular dark field images performed by an enhanced machine learning workflow that has been successfully integrated as a plugin into the Nion Swift software platform used to operate the Nion UltraSTEM 100 microscope.

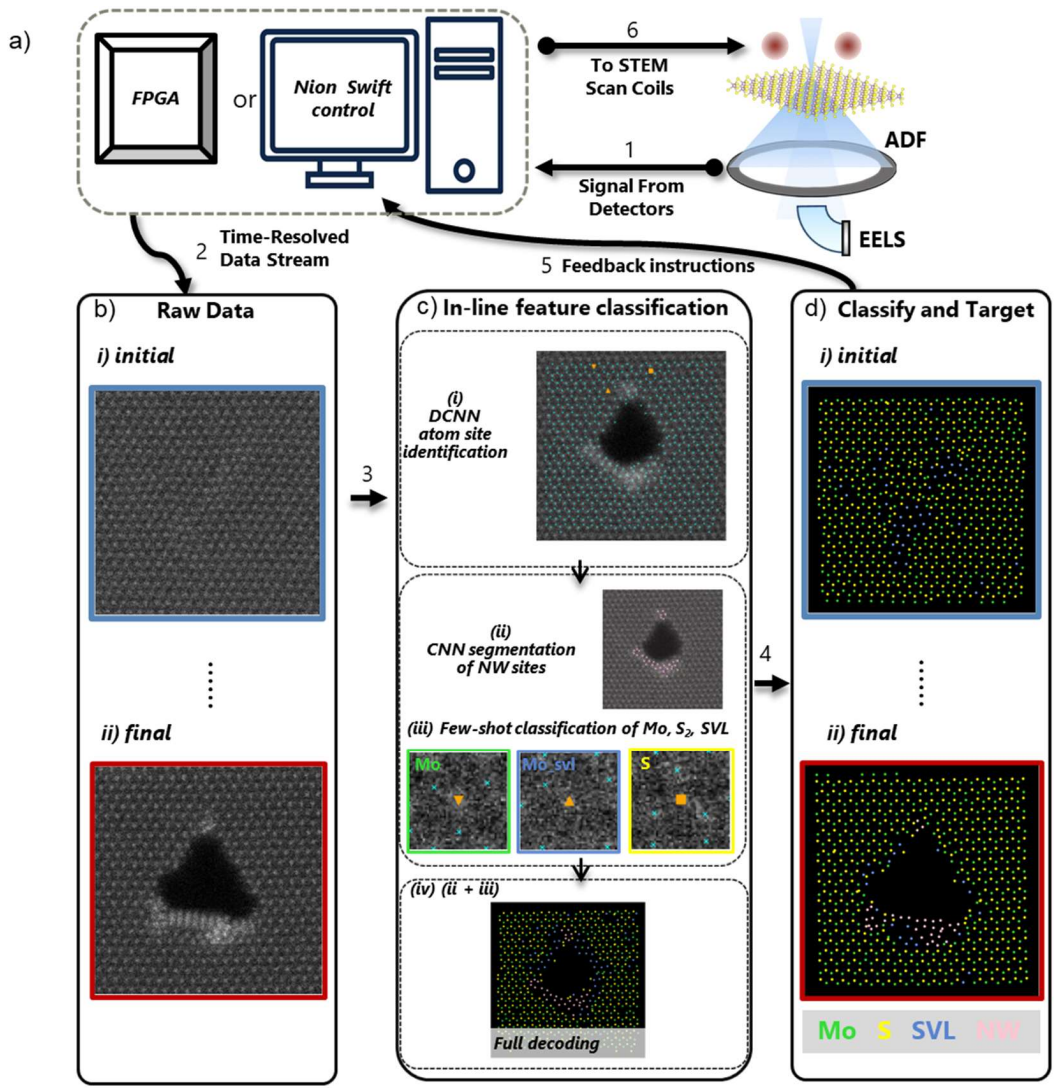

**Figure 1:** Experimental workflow of the Automated Fabrication Controller (AFC) on the Nion UltraSTEM microscopes: **(a)** Signal from the ADF detectors on the microscope are fed into atom identification models through either the Nion Swift controller or our FPGA (1) which is shown in **(b)** as raw data through a live data stream (2). **(c)** The raw data is then run through the machine learning labeling workflow (3) to first **(i)** identify atom positions in the raw image, followed by several post-processes **(ii-iv)** to classify specific atom species. All atom and MoS-NWs are summarized (4) into fully decoded images **(d)**. These decoded sites then provide the targets for the feedback control (5) of the beam positioning system which then sends the positions to the STEM scan coils (6). This process iterates throughout the whole process from the initial to final frame.

Our overall workflow can be broken down into iterative steps that the automated system used as feedback into the beam positioning to appropriately target atoms in the lattice, as seen in Figure 1. At each step, the workflow proceeds as acquiring a HAADF image, decoding the imaging by finding and labeling all the atom sites according to their identities through the machine learning based framework, designating a list of atomic sites for sequential electron beam exposure based on the decoding results, and precisely controlling the beam exposure, driven by either native Nion Swift controller software or a customized FPGA (for more sophisticated scan patterns), to remove these atoms. After either a maximum set time or a maximum set number of atoms have been removed, another HAADF image is acquired to gather new targeting coordinates. In between these steps, drift correction is undertaken in addition to that done during the atom site targeting as described previously. This is done through a quick snapshot being acquired at a larger FOV of 128 nm that is used with an initial snapshot to perform cross-correlation to maintain the same FOV through the initial fabrication steps. Below, we provide detailed descriptions about our machine learning based HAADF image decoding step and our targeted atomic site removal step.

### I. Decoding and labeling atoms through machine-learning based framework

We start with an ADF image of an area of pristine $MoS_2$ monolayer with a square field of view (FOV) size of ~36-100 $nm^2$(6-10 nm × 6-10 nm). The ADF image is fed into our local GPU-equipped desktop for multiple decoding and classification steps (Figure 1c) in which all atom sites identified are classified as one of the following four sites: (1) standard molybdenum (Mo) sites; (2) standard sulfur (S) sites; (3) Mo sites with existing sulfur vacancy, such as single vacancy line, broadly put into a single category (SVL) as sulfur vacancy tend to preferentially form line defects [48, 49]; or (4) outlier sites with significantly altered crystal lattice structure deemed to be existing or potentially forming MoS-NW, broadly named as a single category called nanowire (NW). Details on the machine learning labeling framework are discussed in the following paragraph.

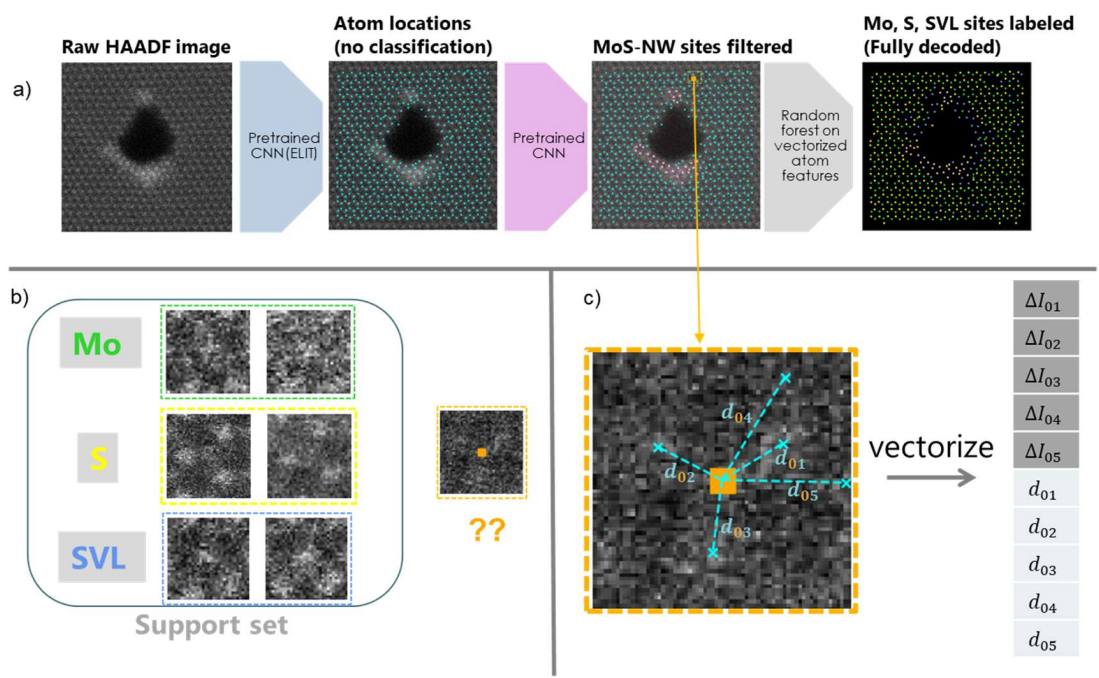

**Figure 2:** Schematics for details of the machine learning aided feature classification models (described in the Methods section): **(a)** Overview of the classification workflow; **(b)** Example instances of the support set curated for Mo, S, and SVL atom sites and their neighborhoods; **(c)** Vectorization scheme, where the local surroundings around an atom site is described by the pixel intensity difference ($\Delta I$) and distance between the atom site and its 5 nearest neighbors.

The machine learning (ML) feature classification framework consists of three parts: an ensemble of Convolutional Neural Network (CNN) models to determine the locations of atom sites from HAADF images acquired from the microscope (Figure 1c-i), a second CNN model to segment nanowire MoS-NW regions from the HAADF images (Figure 1c-ii), and a random forest classification model to classify atom sites into Mo, S, or SVL sites (Figure 1c-iii). A sequential flow diagram for this workflow expanded from Figure 1c is shown in Figure 2a. The output of the framework is fully labeled maps of all atom locations and classifications used to determine beam exposure targets in the next iterations. We note that while S and Mo can potentially be classified by e.g. a simple pixel intensity threshold, the random forest classifier more robustly handles the multi-class classification task between Mo, S, SVL sites, with the difference between SVL sites and regular Mo sites especially subtle and challenging to heuristically define. An additional byproduct benefit of a random forest classifier is an inherent estimation of uncertainty in prediction. In Figure 3, we present an example of the output of our classification workflow as a fully labeled ADF image (Figure 3b), accompanied with the "uncertainty" quantification of the random forest model prediction, calculated as the averaged class probabilities predicted by all the trees in the random forest (Figure 3c-3e). In Figure 3f, we plot the distance between all the Mo atoms, including (regular) Mo sites and SVL sites, and their 1st, 2nd, and 3rd nearest neighbors against the class probabilities of SVL. The red line indicates the "ideal" distance between the Mo sites and the nearest S sites in a pristine $MoS_2$ lattice (note that this is not the Mo-S bond distance, but rather the distance between the Mo atom and the midpoint of the S-S bond, as HAADF images represent a 2D projection of the 3D lattice). Those atoms predicted as regular Mo (low SVL class probabilities) tend to be surrounded by intact crystal lattice, with distances to all three nearest neighbors close to ideal Mo-S value, while the atoms predicted with high SVL class probabilities have a higher likelihood of distorted nearest neighbor distances. However, as one would expect in a complex atomic environment, none of these three nearest neighbor distances alone would be sufficient to define a reliable threshold to identify regular Mo vs. SVL.

On our local computer equipped with an A100 GPU, this entire workflow can generally be completed within ~1 s, with beam blanked in the duration to minimize unnecessary beam exposure. We refer the readers to the Methods section for more details about this machine learning framework.

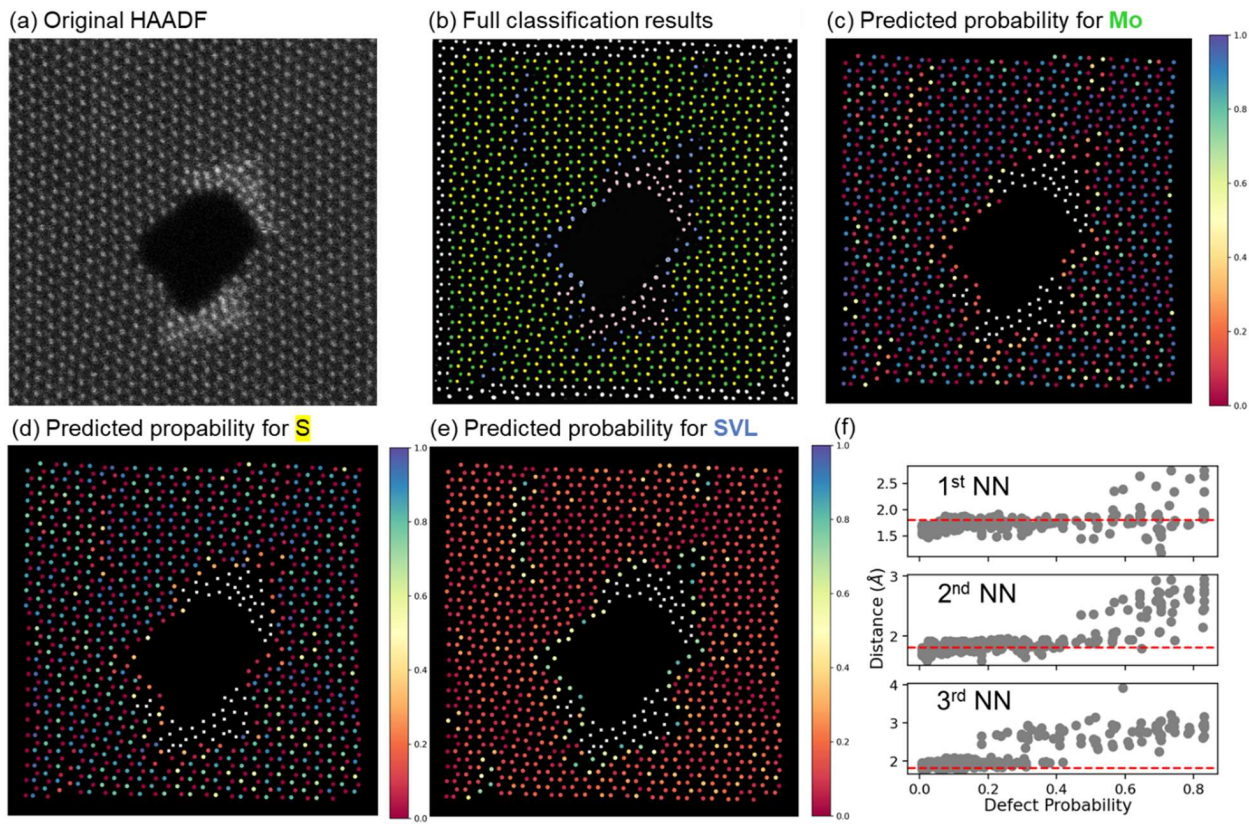

**Figure 3.** Uncertainty quantification of random forest classification between Mo, S, and SVL sites. (a-b) an example of (a) raw HAADF and (b) fully labeled image at a certain iteration in our fabrication workflow, with green, yellow, cyan, and pink representing Mo, S, SVL, and NW sites. (c-e) Predicted probability of the atom belong to (c) Mo, (d) S, and (e) SVL site by "voting" of trees in the random forest classifier. (f) For all the atoms classified as Mo or SVL, the distance between the atom and its 1$^{st}$, 2$^{nd}$, and 3$^{rd}$ nearest neighbors vs. the predicted class probability of SVL. Red lines indicate expected Mo-S$_2$ distance (~0.18 nm) in an ideal MoS$_2$ lattice.

## II. Targeted removal of sulfur sites to fabricate desired MoS-NW structure

With accurate atom labeling throughout machine learning based decoding framework and precise atom site targeting made possible through standard "point and shoot" methods for removing singular (sulfur) atoms from the monolayer and altering the MoS$_2$ crystal structure, a robust experimental workflow was developed to reliably fabricate the nanopores with the MoS-NW edge structures. To allow for additional flexibility of the control of the beam path, an external scan control system driven by our customized FPGA [50] was also utilized. In previous studies, the beam path for each target site was constant in time, i.e., a point dwell. With the native Nion controller this could be extended to small rectangular raster scans centered at target sites, but this results in an unavoidable asymmetric dose profile because of the well-known flyback effect. Using our customized FPGA scan controller, however, a much larger variety of scan paths are possible; here we demonstrate using an Archimedean spiral scan of size 2-5 Å such that the dose was concentrated in the center and radially symmetric across the spiral scan profile. Using the advanced targeting system the initial sulfur sites were targeted using a small triangular mask in the center of the field of view. It was found that, due to the preferential removal of the sulfur atoms from the $MoS_2$ lattice, the sulfur targeting was found to be more reliable to the formation of the MoS-NW than attempting to move the Mo atoms. When targeting the Mo atom sites, vacancy formation was not able to be controlled as easily due to seemingly random formation of neighboring S vacancies next to the targeted Mo sites, making controlled defect fabrication more difficult. However, when S sites were targeted along crystal axes, vacancy formation was relatively controllable. The preferential displacement of S we observe agrees with previous works pointing out that S displacement can occur relatively easily, even well below knock-on threshold if localized electronic excitations preexists, while Mo has significantly larger displacement threshold ($E_d$) than S.[51, 52]

The workflow always starts with the formation of a nanopore on a pristine $MoS_2$ monolayer at the center of the field of view (FOV) through sequential localized beam exposure on several sulfur sites for 1-2 second each at the center of the field of view (FOV) within a triangular mask, a.k.a. "blasting" the sulfur atoms. (Figure 4a-d) Within $MoS_2$ these nanopores have been shown to have interesting functional properties as they are commonly found to be decorated by the MoS-NW structure which introduces a metallic edge structure connected to the semiconducting bulk monolayer, separated by a commonly found insulating on-lattice edge structure.[17] In our workflow, these nanopores also effectively serve as "seeds" for MoS-NW structures to form on the edges. After the nanopore is stabilized and grows in size and defects continue to be generated around the nanopore, the number of MoS-NW sites (pink in Figures 1 and 2) identified by the ML models concurrently start to increase and can be associated with the pileups of Mo atoms around the nanopores. After another one or two iterations, the Mo pileups can order into the desired MoS-NW edge structures. Thus, controlling the fabrication of these MoS-NW structures needs robust monitoring of the occurrence and growth of nanopore and MoS-NW sites in the vicinity of the nanopore, as well as carefully designed strategies according to different stages of growth and intended final MoS-NW structure to be achieved. When the number of MoS-NW sites detected by our machine learning framework exceeds a certain threshold, the fabrication strategy diverges depending on the desired final MoS-NW structure. The seeded nanopore can be targeted and expanded with the goal of forming more continuous MoS-NW structures along the edges of the nanopore, a fabrication strategy we refer to as "targeted NW growth". In these cases, we continue to focus on the center of the FOV but replace the initial triangular mask with a larger triangular mask to form more sulfur vacancies and free Mo atoms to form the MoS-NW on the desired edge of the nanopore. An example of the MoS-NW growth along an edge of the triangular mask can be seen in Figure 5a. In addition, various other control parameters such as additional masks, stage shifts or rotations can be implemented to make more advanced structures depending on the pre-defined parameters. Using this procedure, a variety of structures, such as freestanding MoS-NWs, as well as more complex structures, such as truss-like lattices of MoS-NWs, can be reliably formed within the $MoS_2$ monolayer. In Figure 5b, a different growth strategy is applied, and the parameters call for the formation of a freestanding MoS-NW between two nanopores. This task is accomplished using a second triangle mask at the mirror image location and orientation of the original nanopore within a larger FOV to form another seed nanopore, forming the MoS-NW along the sandwiched edge. As the two nanopores expand towards each other, targeting selective atom positions along the edge between the nanopores introduces this freestanding MoS-NW without introducing strain into the surrounding lattice. This larger FOV requires a stable network to decode the image with different pixel to angstrom ratios and spatial resolutions to reliably form these larger structures.

In a third possible MoS-NW growth strategy, we also attempt to grow the MoS-NW structures into 1D-2D heterostructures in a controllable and directional manner that spans across the entire FOV. The strategy to grow 1D-2D MoS-NWs that expands directionally is referred to as "directional growth" and outlined in Figure 4e-h. With our current seeding method, multiple MoS-NW clusters tend to be generated stochastically around the seed nanopore, so the first step in this "directional growth" strategy is to select a group of MoS-NW sites to grow upon. Once a significant number of MoS-NW sites have been detected, instead of the application of a larger triangular mask as in the "targeted growth" strategy, a DBSCAN [53] is done across all the MoS-NW sites to separate all MoS-NW sites into clusters. The largest cluster is then chosen as the candidate to be grown. As we have not yet implemented any robust temporal tracking algorithms to follow the movement of atoms across iterations and each HAADF acquisition is treated as an independent image, the largest seeded MoS-NW cluster is chosen to be grown under the assumption that the largest cluster will continue to grow and remain the largest cluster, thereby ensuring the same MoS-NW cluster is being targeted throughout the iterations. In practice, we found this strategy to work relatively well, with the algorithm sometimes fluctuating between several small MoS-NW clusters in the early phases but steadily locking onto growing the largest cluster once such a MoS-NW cluster emerges after several iterations. Once the largest MoS-NW cluster is identified, the NW sites in the cluster is fitted to a line segment, ~1 nm extensions are drawn out from both ends of the line segment, and sulfur sites sufficiently close (we currently found a good threshold to be ~0.2-0.3 nm) to the two extended segments are targeted for blasting in the next iteration. An example of such line fitting and blasting site designation is shown in Figure 4h. In Figure 5c, the process of fabrication a long MoS-NW spanning across the entire FOV using the method described above is shown.

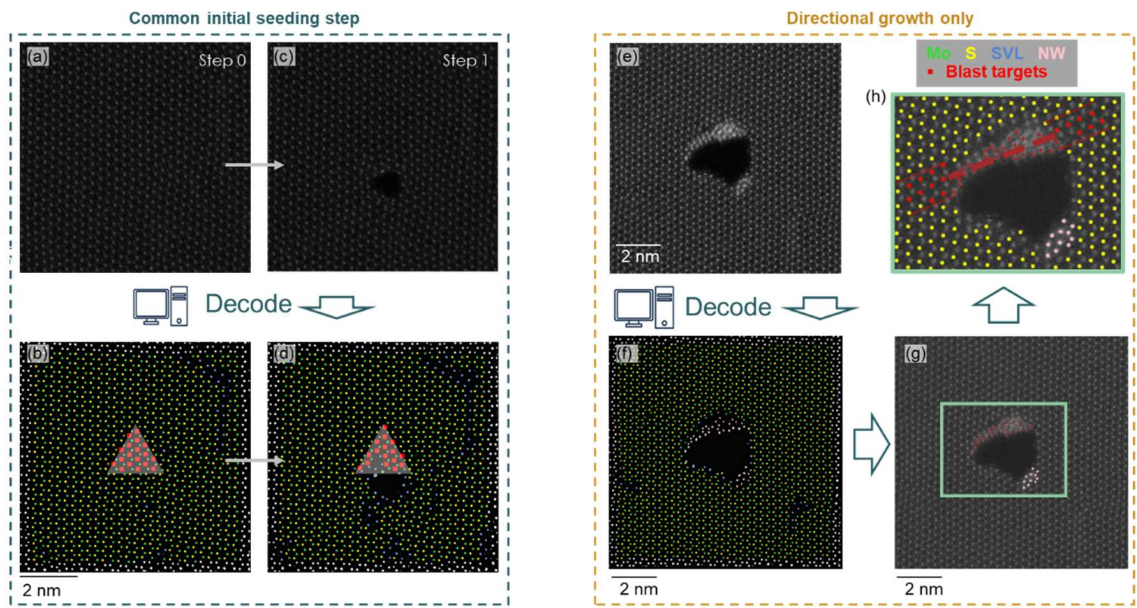

**Figure 4:** MoS-NW site identification and selection: (a-d) Initial seeding of nanopore from pristine $MoS_2$ monolayer, regardless of the desired MoS-NW structure to be fabricated, with (a,b) first attempt to eject S within the center triangular mask and (c,d) continuing enlargement of the pore after it is seeded but MoS-NW has yet to form. (e-h) Additional steps implemented only for directional growth of MoS-NW. with the raw ADF acquisition (e) decoded in (f), largest cluster selected in (g), and the MoS-NW sites in the largest cluster fitted to a line segment and sulfur sites on the two ends of the line segment chosen for targeted removal in (h).

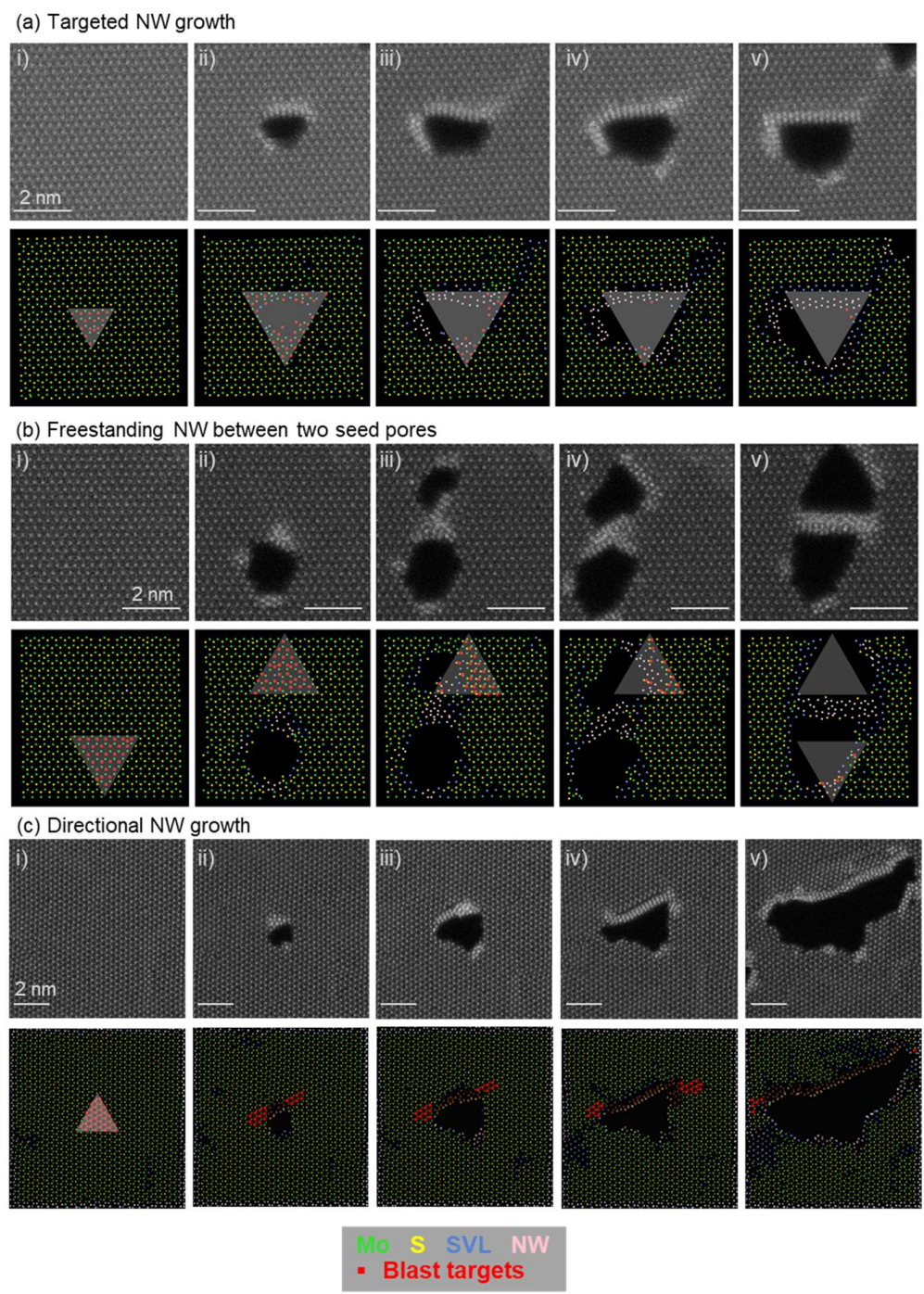

**Figure 5:** A selection of results from different MoS-NW fabrication strategies: (a) Example of the technique used to form a MoS-NW along a pre-selected edge of a nanopore using a seed hole and then preferential S-site targeting. (b) Example of the technique to make a freestanding MoS-NW using two triangle masks. (c) Example of the technique to first form a MoS-NW on a desired edge and then grow along crystallographic directions. In all three subfigures, *(i-v)* represent raw ADF acquisition and decoded images with beam targets for the next iteration overlaid, from the beginning to the end of the iterative workflow. Mo (green), S (yellow), SVL (blue), and NW (pink) sites are labeled according to their decoded identity. The targeted S sites are labeled red. All snapshots and decoded images scalebars are 2nm.

**Conclusions:**

In this work, we presented our explorations in fully autonomous fabrications of atomic-level defects using electron beams in STEM. As an example, MoS-NW structures were successfully fabricated in a controllable and reliable manner by iterative beam exposure to remove sulfur atoms from $MoS_2$ and HAADF acquisition in between to monitor the morphological change of the sample. A comprehensive machine learning (ML) framework was deployed to decode the HAADF acquisition and fully label the acquired images with identities of all atoms. As evident in the multiple series of results presented in this article (e.g., Figure 5), the ML framework consistently achieved reliable atom classification at various pixel sizes, targeted fabrication morphologies, and stages during the fabrication process despite being trained on a small set of labeled data (~20 images with labeled MoS-NW regions and ~40 labeled atoms with their local surroundings). The decoded atomic identities output by the ML framework are then fed into an autonomous decision-making platform with pre-defined fabrication strategies to selectively remove sulfur (S) sites. Finally, coordinates of the targeted sulfur sites are sent to scan control to execute sequential beam exposure, with our FPGA-controlled high-precision custom scan routine incorporated to provide full flexibility in scan path and duration. Altogether, this work demonstrates a concerted development and implementation that combines machine learning and automated algorithms with programmable and flexible electron beam control enabled by our customized FPGA controlling module to realize autonomous feature characterization and fabrication on nanoscale. Not only do the fabricated MoS-NW structures represent the potential of unique mechanical and electrical possibilities, but the autonomous platform we present here also can be adapted to other material or sample and can be extended to other 2D materials for other defects and heterostructures beyond $MoS_2$.

The autonomous platform presented in this work also represents an exciting starting point for future development to achieve higher levels of intelligence, flexibility, and insight during the autonomous procedure. On the image decoding side, in this work reliable atom classification in ADF images is achieved by a combination of relatively simplistic, small-data machine learning algorithms and intuition-based embedding of atomic neighborhood. Properly labeled atomic features include fundamental identities such as the Mo and S sites in regular $MoS_2$ lattice as well as simple defects such as MoS-NWs or SVLs. While the results are promising, one can fathom scenarios where more sophisticated models/algorithms might be needed for complex structural identification in other use cases. Due to the inherent difficulty in acquiring large amounts of labeled data in materials science regime, it would be interesting to always explore the limitation in capability of simpler machine learning methods that can be trained with less data and incrementally move to "larger" models with more trainable parameters when the simpler models start to fail. Theory and simulations can also help provide training data to establish foundation models or even physical insights that can be directly infused into the model. More information, such as crystallographic information or even *in-silico* production of defects, can be extracted from more deliberately designed simulation routines and improve the robustness of the model to be transferred to real-world experimental scenario. In addition, temporal connections between HAADF acquisitions throughout the workflow can track the motion of atoms and evolution of defects to guide more intelligent decision making that connects early, subtle atom displacements to subsequent defect formation on larger length scales. Developments on both the software and hardware sides are also desirable for strategic control of electron dose to minimize unwanted beam effects on the 2D materials from both the imaging stage and the targeted blasting stage. Finally, merging these workflows with AI-agents that can have dynamically changing rewards that are altered by the human operator can provide a template for atomic manipulation agents that continually learn and adapt to changing microscope conditions, as has been recently demonstrated in some microscopy experiments.[28, 54]

**Methods:**

Sample preparation:

The MoS$_2$ monolayers were synthesized using chemical vapor deposition in a 2 in. diameter quartz tube at a growth temperature of 750 °C using a hinged tube furnace (Lindberg/Blue). The growth substrates (SiO$_2$ (300 nm)/Si) were first cleaned with isopropanol and then spin-coated with perylene-3,4,9,10-tetracarboxylic acid tetrapotassium salt before being dried and placed face-down above an alumina crucible containing MoO$_3$ powder (∼5 mg) (99.9%, Sigma-Aldrich). At this point the crucible with the substrate was then inserted into the center of the quartz tube within the furnace. Another crucible was filled with S powder (∼0.7 g) (99.998%, Sigma-Aldrich) and inserted at a region located 20 cm upstream from the other crucible in the quartz tube. This location was found to be an area with a temperature around 180 °C when the furnace is heated to 750 °C. To perform the CVD, the tube was then evacuated to ∼5 mTorr where upon a 70 sccm (standard cubic centimeters per minute) flow of Ar at atmospheric pressure was established. At this point the furnace was then heated to 750 °C at a rate of 30 °C/min and held at temperature for 4−6 min. The furnace was then allowed to naturally cool down to room temperature.

The newly grown MoS$_2$ monolayers from this process are then transferred from the SiO$_2$/Si substrate onto gold mesh holey carbon Quantifoil TEM grids using a wet transfer process for STEM imaging. PMMA (Poly(methyl methacrylate)) was spun coated onto the SiO$_2$/Si substrate with monolayer MoS$_2$ crystals on the surface at 500 rpm for 15 s followed by 3000 rpm for 50 s. The PMMA-coated sample was then placed into a 1 M KOH solution to remove the SiO$_2$ substrate. This leaves the PMMA film with the MoS$_2$ monolayer on the solution surface. The KOH residue was then rinsed from the sample film with DI water. A TEM grid was then placed into a funnel filled with DI water. Using a glass slide the washed PMMA/MoS$_2$ film was then transferred to float on the DI water. As the water drains the film is then placed onto the TEM grid. The samples are then baked to dry at 80 °C before being placed into acetone to soak for 12 hours to remove remaining PMMA before a final rinse in IPA, leaving a clean monolayer surface. Prior to STEM experiments, the specimens were baked at 160°C in vacuum overnight to reduce surface contamination.

STEM imaging:

An aberration corrected Nion UltraSTEM100 electron microscope was used for all the STEM imaging experiments while operating at 60kV. The custom-built FPGA control system described in this study was based on a National Instruments DAQ (PXIe-6124) and field-programmable gate array (FPGA) system (PXIe-7856R) in a PXIe-1073 chassis that was interfaced with the beam control system of the Nion UltraSTEM100. Input coordinates from customizable MATLAB code were then put into a LabView program that was used to control the STEM scan coils to position the focused STEM probe along well-defined scanning pathway shapes. The custom scan control system has a maximum IO rate of 2M Samples/s.

Machine learning models and training:

The code and data used to train the machine learning models (ELIT ensemble for atom location determination, random forest classification model for classification of Mo, S, and SVL sites, and the U-net model for MoS-NW segmentation) are made available in our open-source repository at https://github.com/zijiewu3/MoS_NW_AFC. Several key details are highlighted below.

The initial CNN model for identification of atom sites is directly based on our recent work on Ensemble-Learning Iterative Training (ELIT) model for atom identification in STEM.[35] The ELIT ensemble is pretrained on multislice simulation of $MoS_2$ crystals, with the best model chosen and augmented based on small amount of experimental data. We refer the interested readers to that work for the details on this model.

Built upon the known atom site positions from the ELIT model, a random forest classification model is trained using Scikit-learn[55]. Hyperparameters of the random forest model are tuned through random search based on a 5-fold cross validation of the supporting set. A support set of 55 hand-labeled instances (21 Mo, 24 S, and 10 SVL sites) from previously acquired HAADF images of $MoS_2$ is used to train the random forest, each instance describing the local neighborhood of a (regular) Mo, S, or Mo SVL atom. Some examples of the atom sites we collected in the support set are shown in Figure 2b. We believe 55 instances represent a reasonable number of training instances that a domain expert can quite easily collect and label with at most several hours of work. While we expect even more instances in the support set will further improve the model performance, our results in the later section show that this relatively modest support set already performs sufficiently well, demonstrating the significance of our choice and design of the models. The model learns on an embedded representation of the local environment of each atom. Multiple embedding algorithms can potentially be deployed here (e.g. analytical atomic neighborhood descriptors such as SOAP [56] and ACE [57]), but we found that somewhat surprisingly, a seemingly simple vectorization of the distance and intensity of the atom's five nearest atoms (Figure 2c), while lacking permutation invariance, already provides sufficient performance for our purpose of encoding atom's local environment. Since this atom classification model relies on information about the relative distances and intensity differences between nearby atom sites, its performance depends on the ELIT model accurately locating these atom sites in the first step.

In the case of MoS-NW, due to the collapsing of the crystal lattice, the atom site location can become random and/or ambiguous, hindering the random forest classification model from learning useful information from the data. So, we instead deploy another CNN model to directly segment MoS-NW regions. The CNN model uses a lightweight U-net [58] architecture built with AtomAI [59] (which uses a PyTorch[60] backend) and is trained from ~30 labeled $MoS_2$ HAADF images. All input images are first rescaled to a uniform pixel dimension of 0.085 Å/pixel. Random transformations, including horizontal and vertical flip, rotations, cropping, and color jittering, are performed on images in the training set for data augmentation. The model is then trained with Adam optimizer and a mixed loss function of dice loss and Intersect over Union (IoU) loss. The CNN MoS-NW model is deployed after the ELIT atom identification model and before the random forest model to classify the Mo, S, SVL sites; atoms in the CNN-segmented NW regions are automatically labeled as NW sites and not fed into the random forest model.

**Author Contributions:**

Z.W., M.G.B., and K.M.R. conducted STEM experiments, constructed machine learning algorithms, and performed data analysis, and lead the writing of the manuscript. K.M.R. and A.G. provided assistance and guidance with the coding and data analysis with additional support in the construction of the machine learning models. S.J. provided experimental assistance with the beam control system during STEM experiments. R.R.U. provided guidance during experiment planning. K.X. provided $MoS_2$ monolayer samples. All authors additionally contributed and approved the final manuscript.

**Acknowledgements:**

All STEM experimental work was supported by the Center for Nanophase Materials Sciences (CNMS), which is a US Department of Energy, Office of Science User Facility at Oak Ridge National Laboratory. Feedback and guidance on machine learning workflow development and data analysis (A.G.) were funded by U.S. Department of Energy, Office of Science, Office of Basic Energy Sciences, Materials Science and Engineering Division. Material synthesis, sample transfer (K.X.), and development of the scan control (S.J.) were supported by the U.S. Department of Energy, Office of Basic Energy Sciences, Division of Materials Sciences and Engineering.